\documentclass[fleqn,usenatbib]{mnras}

\usepackage{newtxtext,newtxmath}

\usepackage[T1]{fontenc}

\DeclareRobustCommand{\VAN}[3]{#2}
\let\VANthebibliography\thebibliography
\def\thebibliography{\DeclareRobustCommand{\VAN}[3]{##3}\VANthebibliography}

\usepackage{graphicx}	
\usepackage{amsmath}	

\title[NEOSSat Observations of Three Hot Jupiters]{NEOSSat Observations of Three Transiting Hot Jupiters}

\author[Fox, Wiegert]{Chris Fox$^{1,2}$\thanks{Contact e-mail: \href{mailto:cfox53@uwo.ca}{cfox53@uwo.ca}}, Paul Wiegert$^{1,2}$
\\
$^{1}$Department of Physics \& Astronomy, The University of Western Ontario, London, Ontario, Canada\\
$^{2}$Institute for Earth and Space Exploration (IESX), The University of Western Ontario, London, Ontario, Canada}

\date{Accepted for publication by the Monthly Notices of the Royal Astronomical Society, September 5 2022}

\pubyear{2022}

\begin{document}
\label{firstpage}
\pagerange{\pageref{firstpage}--\pageref{lastpage}}
\maketitle

\begin{abstract}
The Near Earth Object Surveillance Satellite (NEOSSat) is a Canadian-led 15~cm Earth-orbiting telescope originally designed to detect asteroids near the Sun.  Its design is however also suitable for the observation of exoplanetary transits of bright stars. We used the NEOSSat platform to perform followup observations of several Transiting Exoplanets Survey Satellite (TESS) targets, both as a demonstration of NEOSSat capabilities for exoplanetary science and improve the orbital ephemerides and properties of these exoplanet systems.  We are able to recover / confirm the orbital properties of such targets to within mutual error bars, demonstrating NEOSSat as a useful future contributor to exoplanetary science.
\end{abstract}




\section{Introduction}
The Near Earth Object Surveillance Satellite (NEOSSat) \citep{laurin2008} is a Canadian microsatellite originally designed to detect and track near-Sun asteroids.  About the size of a suitcase, NEOSSat orbits Earth in a Sun-synchronous orbit of approximately 100 minutes.  It carries a 15 cm F/6 telescope with a limiting magnitude of $V=19.5$ for a 100 second exposure.  Its spectral range is 350 nm to 1050 nm,  and has a field of view of 0.86 x 0.86 degrees. These characteristics make it suitable for exoplanet transit studies of sufficiently bright stars.

Here we report on a campaign that used NEOSSat to perform followup observations of three exoplanets from the list of systems observed by the Transiting Exoplanets Survey Satellite (TESS) \citep{ricker2014}. TESS has discovered thousands of new exoplanets, and many of these objects require followup observation to confirm or improve the transit ephemeris and other parameters.

Though we will use the NEOSSat data to determine specific parameters of our target systems, our motivation is primarily to assess the capabilities of NEOSSat itself: we aim to demonstrate that NEOSSat can be a useful platform for exoplanetary science; that it can produce reliable parameters that can be used in conjunction with other data sources to better characterize future exoplanetary candidates.

\section{Data Processing}
The data were obtained from NEOSSat on-orbit operations from March 2021 to June 2022. Specific dates of our observations are reported when each system is discussed (below). Raw FITS images from each NEOSSat observing run were downloaded from their repository at the Canadian Astronomy Data Centre (CADC).

Exposure times for targets were as short as 5 seconds and as long as 20 seconds, with cadences of 10 seconds through 20 seconds.  To avoid photometric saturation, we optimized the exposure times for each target so that the peak pixel count was one-third to one-half the maximum value (pixels are 16 bit, thus a maximum count of 65535).  Observations of a single transit event typically included 300-600 individual observations; the exact number obtained depended on factors such as line-of-sight limitations, whether the satellite is over the South Atlantic Anomaly (SAA, where increased particle flux causes excessive noise in the raw images), and certain orbital and operational constraints (such as desaturation of the reaction wheels after the SAA and re-locking onto the target) result in observations typically being limited to less than one hour.

Once downloaded, we used differential photometry to extract the light curve of each target. This is a two step process, based on cleaning and extraction code designed for the NEOSSat mission by Jason Rowe \citep{roweneo} and modified for our purposes.  First, the cleaning algorithm removed instrumental effects (such as bias, electronic and dark noise) and created a dark-subtracted set of science images.  The second step is the extraction process.  Stars are detected in the image using AstroPy's DAOphot routine \citep{AstroPy2013, AstroPy2018} .  The fields are registered and stacked to create a deep master field, from which the list of stars is generated. Affine transforms are used to adjust the positions of the photometric apertures on each image, to compensate for any spacecraft pointing irregularities. After the photometry was extracted, a principal component analysis (PCA) of the brightest non-target stars was performed to remove systematic instrumental trends. The resulting normalized data are suitable for fitting with model transit light curves, and are discussed below.

\section{Analysis}
With the normalized light curves in hand, the available parameter space was searched in order to find a transit model that best fit the system.  In this effort, two publicly available pieces of software were used.  The transit modelling tool PyTransit \citep{parv2015} was used to create hypothetical model light curves  based on a set of parameters. We adopted the quadratic model of \cite{mandelagol2002}, the parameters of which are: planet radius, central transit time, period, semimajor axis, inclination, and two limb darkening coefficients.  To find the light curve parameters that best fit the observations, the Bayesian analysis tool PyMultinest \citep{buch2014} was used.  Combined, these two pieces of software enabled us to search the parameter space that best fit the observed data, with our likelihood function based on the standard $\chi^{2}$ metric.

To reduce the effects of longer term trends in the data that were not completely removed by the PCA, all observations used for generating the light curve and analysis were limited to within one transit duration of the expected central time (based on the Exoplanet Transit Prediction Service \citep{nep2013}).  That is, if the transit had an expected duration of $D$, then we used observations from a time of $-D$ to $+D$ of the expected central transit time, with the expected transit occurring from $-D/2$ to $+D/2$.  

In addition to computing system parameters, we also searched for Transit Timing Variations (TTVs) of each target.  For each individual transit that showed a clear ingress or egress, we ran a separate Bayesian analysis.  This used the best-fit shape parameters from the entire light curve (derived from all transits), but fit the central transit time for each individual transit.  The resultant best-fit time for each individual transit was compared to the best-fit times computed from the entire light curve (all transits). The difference provides an estimate of the timing variation of a particular transit event.

\subsection{Priors}
To fit the observed data to the parameters, our Bayesian fitting algorithm used seven priors: the planet radius, epoch time, orbital period, semimajor axis, inclination, and two limb darkening coefficients.  Though the precise mechanism of hot Jupiter formation is still being debated, strong tidal dissipation resulting in orbital circularization is expected \citep{RasioFord1996, FabryckyTremaine2007, Chernov2017}. Thus our fits assumed eccentricity was zero for all runs. 

The radius prior ran from 0 to +50\% of the initial value reported by TESS.  The central transit time was arbitrarily chosen to be near the center of one of the transits.  The semimajor axis prior was set to be approximately $\pm$50\% of value of the value provided by the Exoplanet Transit Prediction Service \citep{nep2013}.  Inclination values were allowed to range from 60 to 90 degrees.  The longitude of the ascending node is not needed by PyTransit and was not fitted.  The period prior was given a $\pm$0.25 day from the expected value.  Because data was collected over multiple months and non-continuous, a tight period prior was required to prevent the fitting routine from settling into incorrect periods at integer multiples of the true period.  All priors were uniform priors. 

The model used for limb darkening was the quadratic model \citep{mandelagol2002} as implemented by PyTransit.  The two Limb Darkening Coefficients (LDC) priors encompass $\pm0.05$ from the values provided for the star in question from the ExoPlanet Characterization Toolkit from the Space Telescope Science Institute \citep{ctk2018}.  To create the LDC priors, we varied the effective temperature, surface gravity and metallicity inputs of the Toolkit to the extent of their uncertainties.  The resultant range of outputted LDC values (from one extreme to the other) were used as the range of the prior.  Thus, these LDC priors account for uncertainties in the stellar effective temperature, surface gravity and metallicity.

\begin{table*}
	\centering
	\caption{Priors For Targets}
	\label{tab:allpriors}
	\begin{tabular}{rcccl} 
	    Parameter & WASP-43 & TOI-1516 & TOI-2046 & Units\\
		\hline
		Planet Radius & [0.0, 0.25] & [0.0, 0.2] & [0.0, 0.2] & R$_{s}$ \\
		LDC 1 & [0.65, 0.75] & [0.25, 0.35] & [0.35, 0.45] &  \\
		LDC 2 & [0.0, 0.1] & [0.3, 0.4] & [0.2, 0.3] &  \\
		Central Time & [0.0,1.0]+2459695.5 & [0.0,1.0]+2459391.8 & [0.0,1.0]+2459423.5 & BJD \\
		Period & [0.6,1.0] & [1.5, 2.5] & [1.0, 2.0] & days \\
		Semimajor Axis & [3.0, 6.0] & [2.0, 10.0] & [2.0, 7.0] & R$_{s}$ \\
		Inclination & [60.0, 90.0] & [60.0, 90.0] & [60.0, 90.0] & deg \\
	\end{tabular}
\end{table*}

\section{Targets}
Of the three targets analyzed here, two (TOI-1516.01 and TOI-2046.01) were classified as "planetary candidates" by the NASA Exoplanet Archive at the time of observation; they were recently confirmed by \citep{kabath2022}.  Our third target was WASP-43b which is a previously "confirmed" hot Jupiter.

\subsection{WASP-43}
Also known as TOI-656.01, WASP-43b is a well-studied ultra-hot Jupiter (\citet{hellier2011, esposito2017, patel2022}, among others).  The host has an apparent visual magnitude $V=12.3$ and is a K7 main sequence star \citep{hellier2011}.  The orbital period of the planet is 0.813 days.  NEOSSat observations were taken on February 27, April 26, and May 24 of 2022.  The exposure time was 20 second with cadence also of 20 seconds.  The priors used for this target are shown in table~\ref{tab:allpriors}.  
\begin{figure*}
 \includegraphics[scale=0.30]{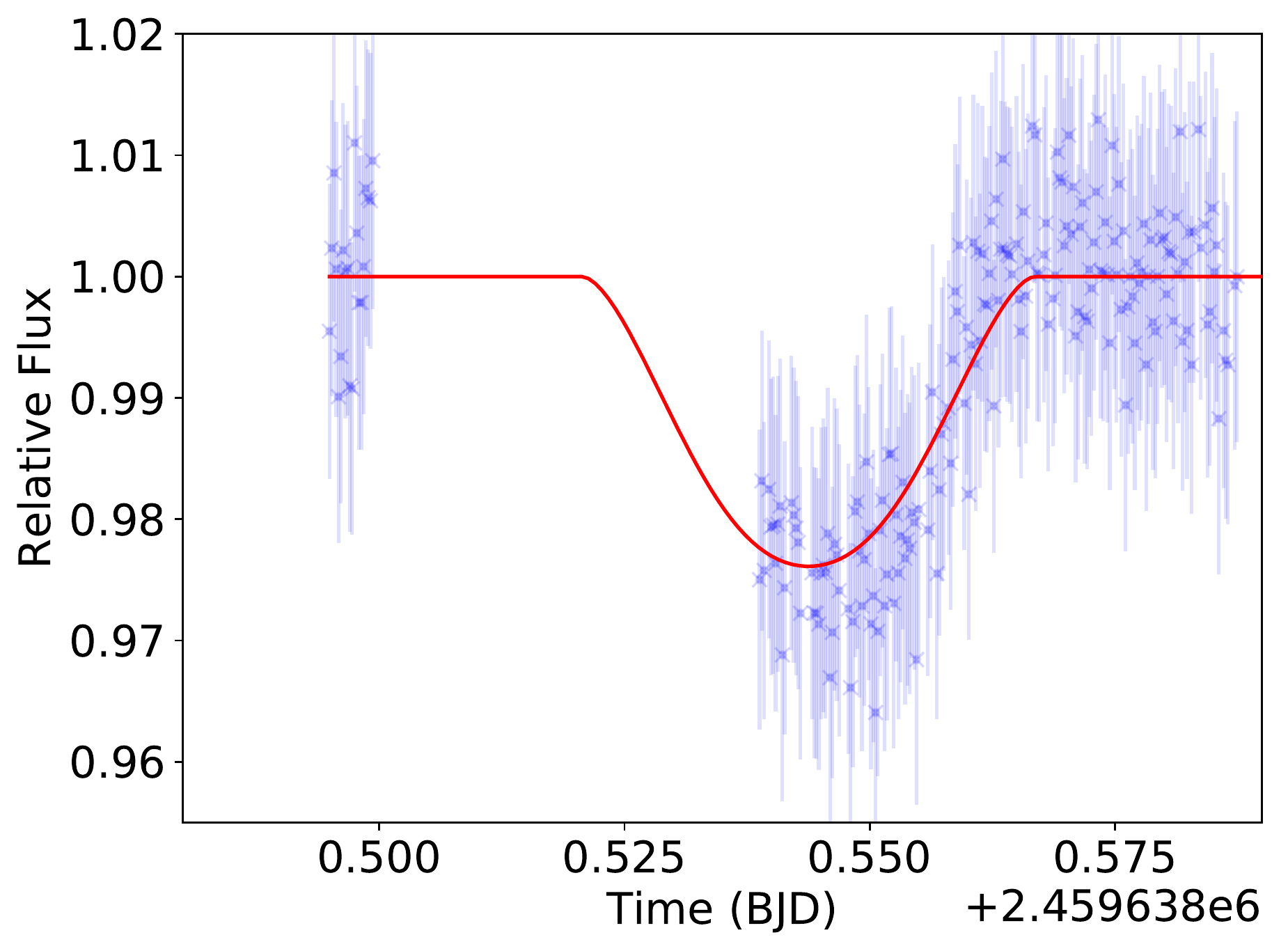}
 \includegraphics[scale=0.30]{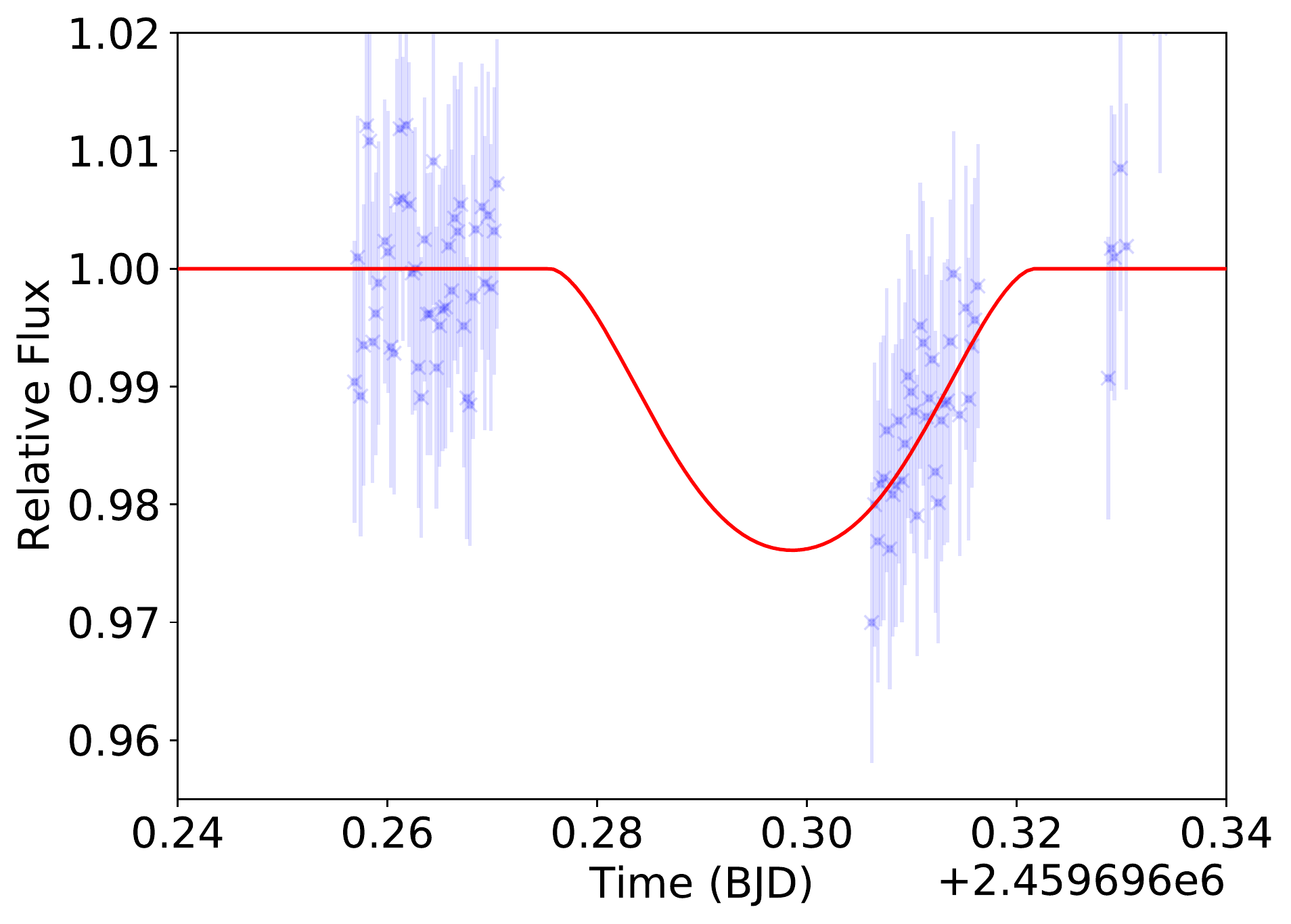}
 \includegraphics[scale=0.30]{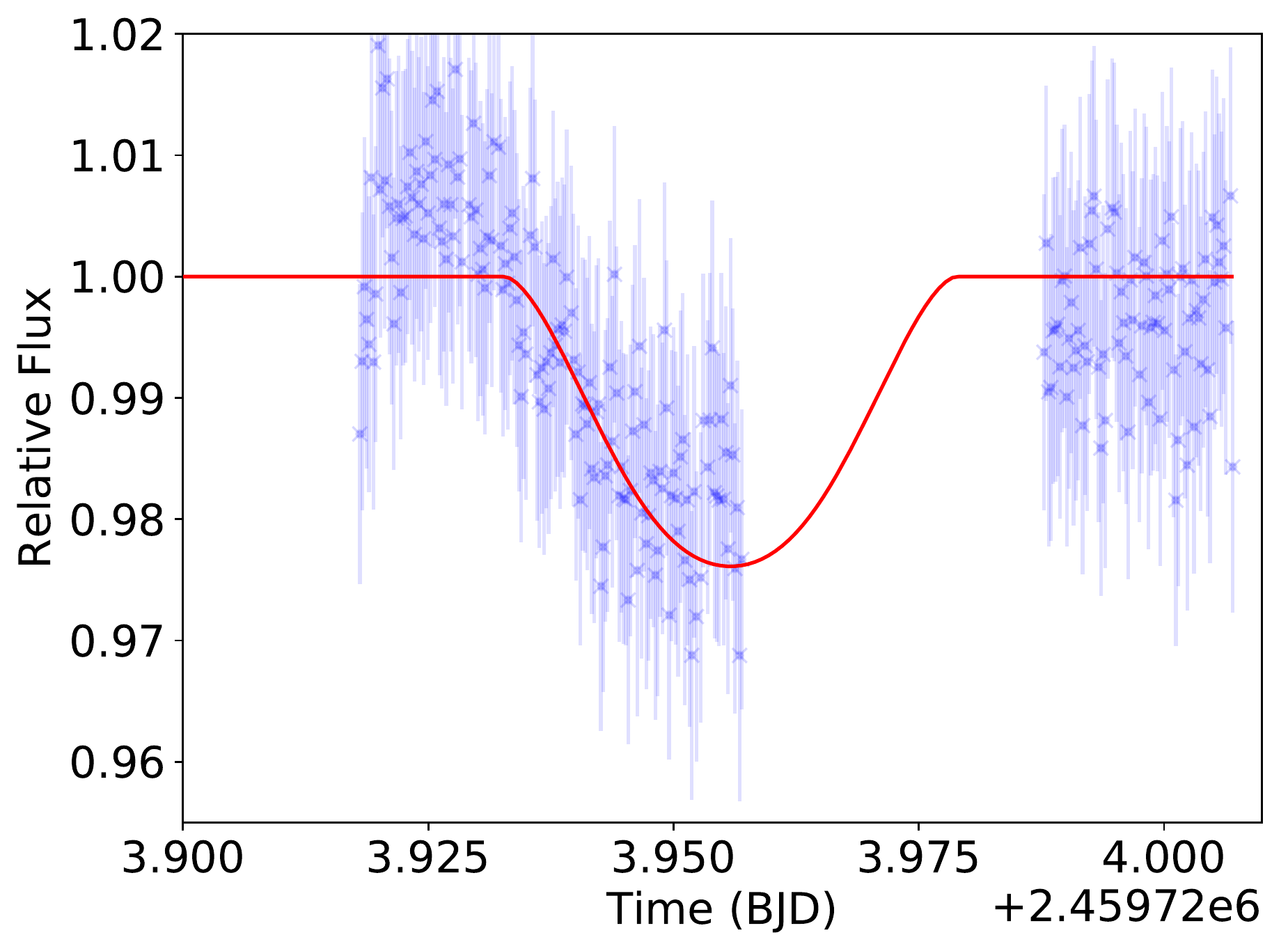}
 \caption{Normalized flux observations of of WASP-43b.  The red line is our fitted light curve.\label{fig:toi656zoom} }
\end{figure*}

\begin{figure*}
 \includegraphics[scale=0.75]{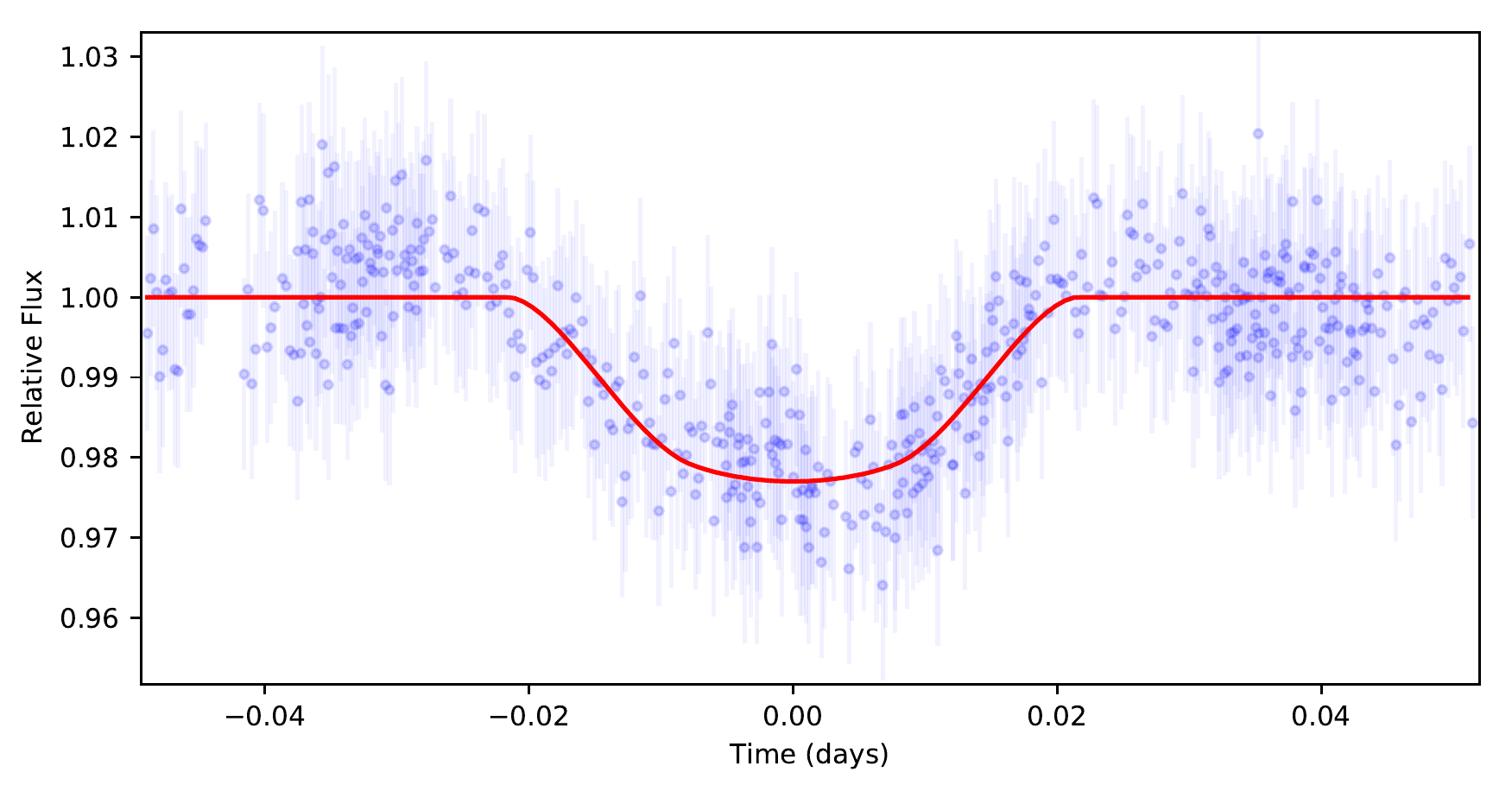}
 \hspace*{-0.29cm}\includegraphics[scale=0.75]{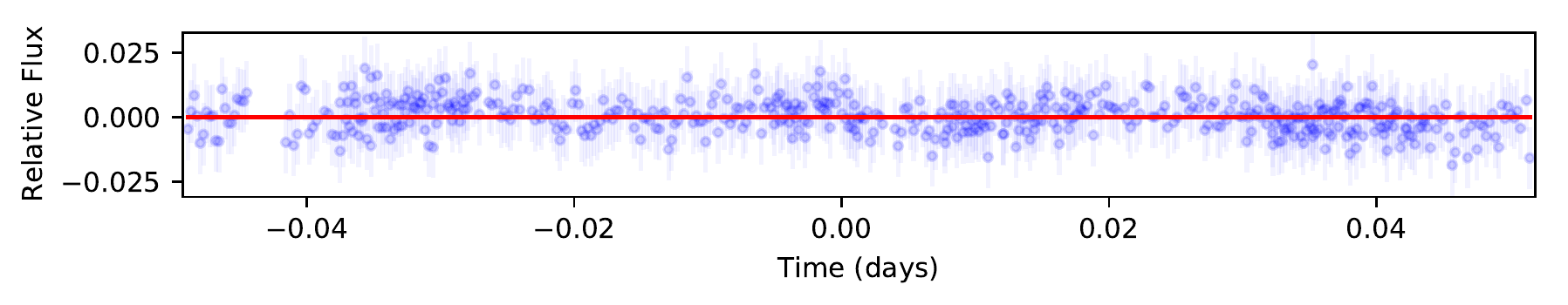}
 \caption{Phase-folded light curve and residuals of WASP-43b. The red line is the fitted light curve.\label{fig:toi656all} }
\end{figure*}

Individual transit event data for WASP-43b are shown in figure \ref{fig:toi656zoom}, and the phase-folded light curve is shown in figure \ref{fig:toi656all}.   The NEOSSat results for WASP-43b are largely consistent with other results \citep{hellier2011, esposito2017}, and are presented in table~\ref{tab:allposts}.  Our results suggest a slightly higher impact parameter than other sources \citep{esposito2017, patel2022}.  Because of the longer cadence and fewer observed transits, this target has the highest error among our three targets, but are within the mutual error bars of other sources.  Of the three transits, one has excellent coverage of egress, and the other of ingress.  Using these, we find all TTVs for observed events to be consistent with 0 $\pm30$ seconds.

\subsection{TOI-1516}
TOI-1516.01 is a planetary candidate observed by TESS, observed in FFI (full-frame-image) in Sectors 17 and 18.  The host star is a $V=10$ F8 main sequence star \citep{kabath2022}.  NEOSSat observations were taken on March 28, May 14, May 18, June 26, September 2 and October 9 of 2021.  The exposure time was 5 seconds with a 10 second cadence.  This planet was validated by \citet{kabath2022} after the NEOSSat observations were taken, but before the writing of this paper.  The priors used for this target are shown in table~\ref{tab:allpriors}.
\begin{figure*}
 \includegraphics[scale=0.30]{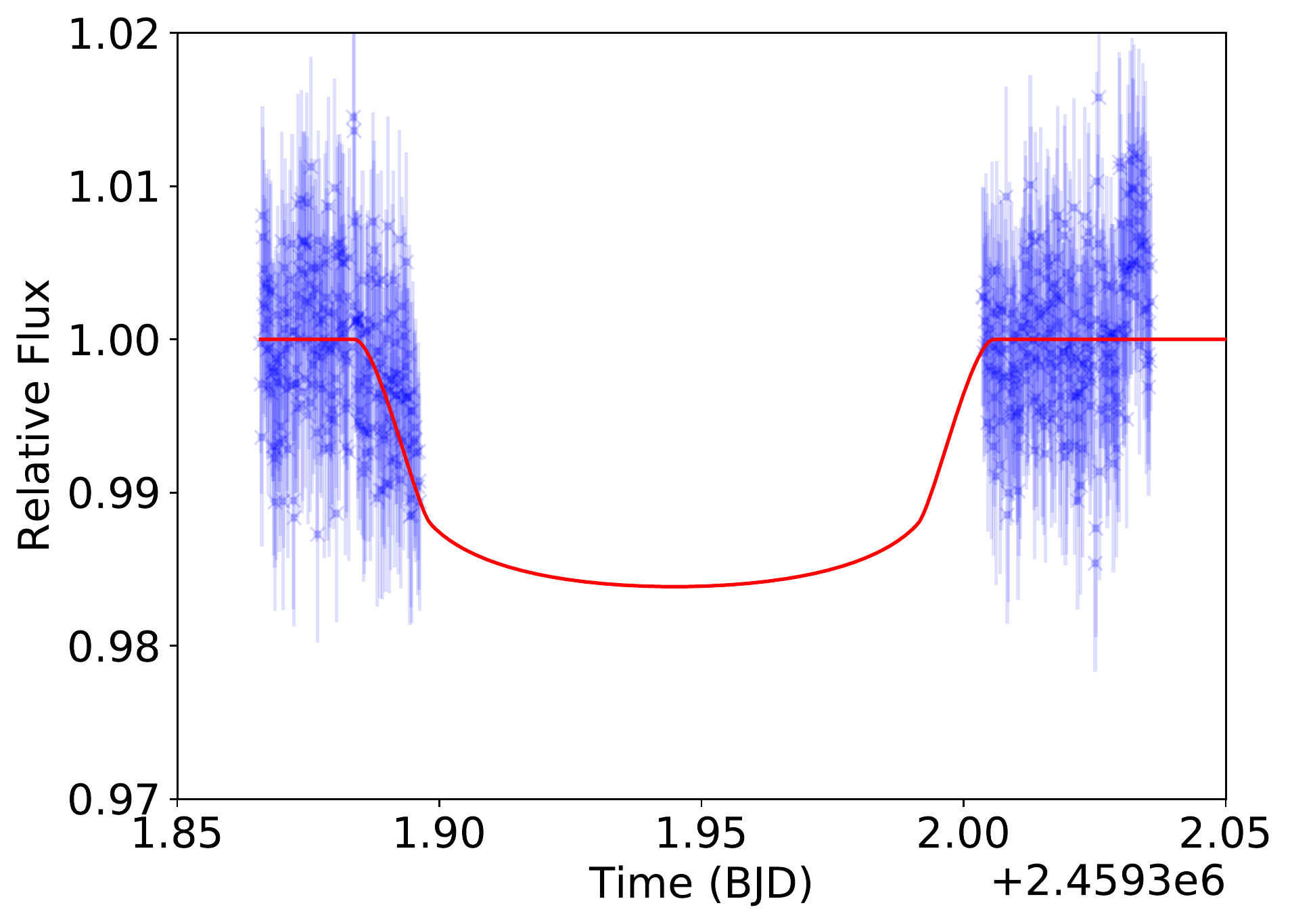}
 \includegraphics[scale=0.30]{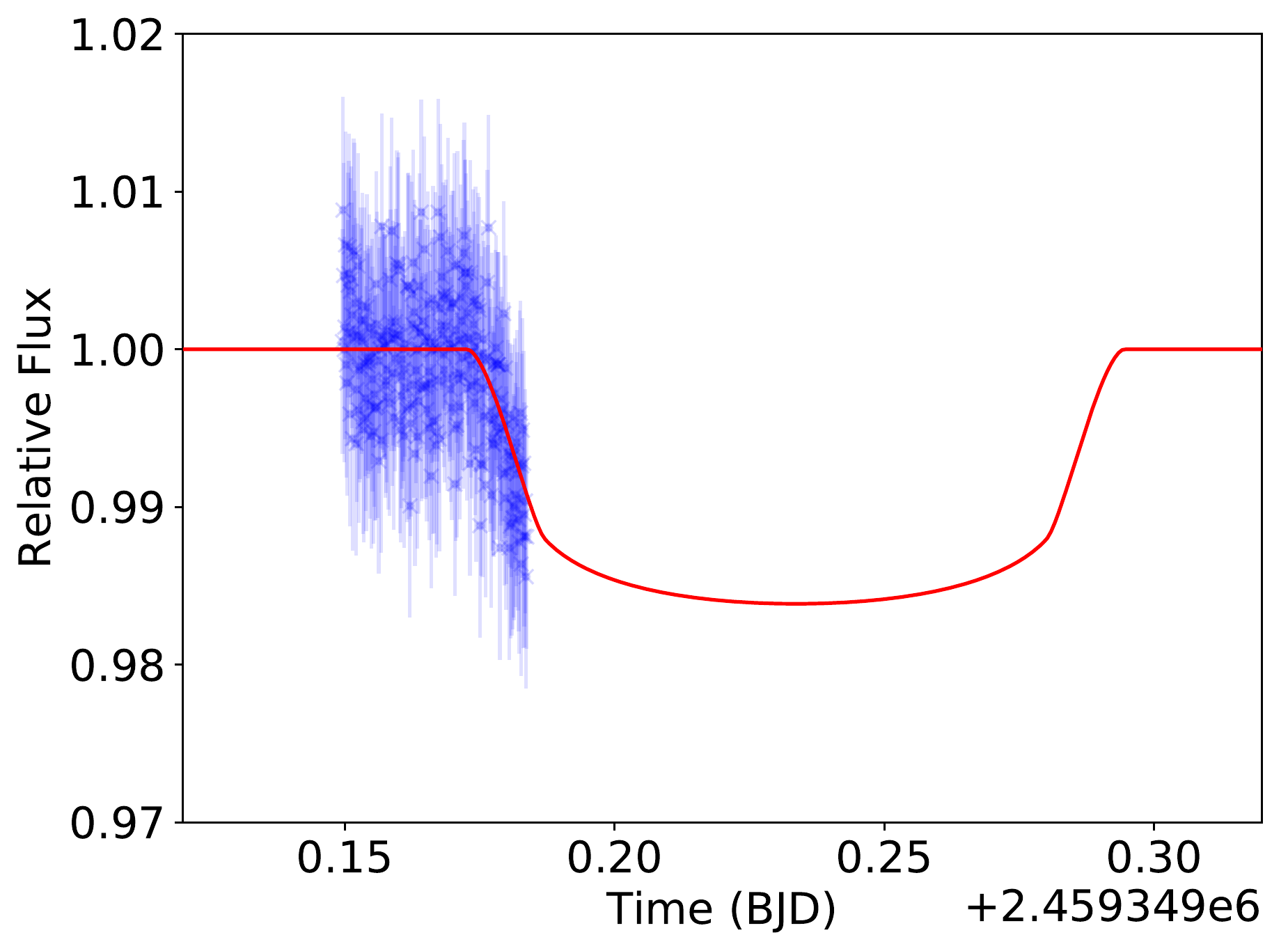}
 \includegraphics[scale=0.30]{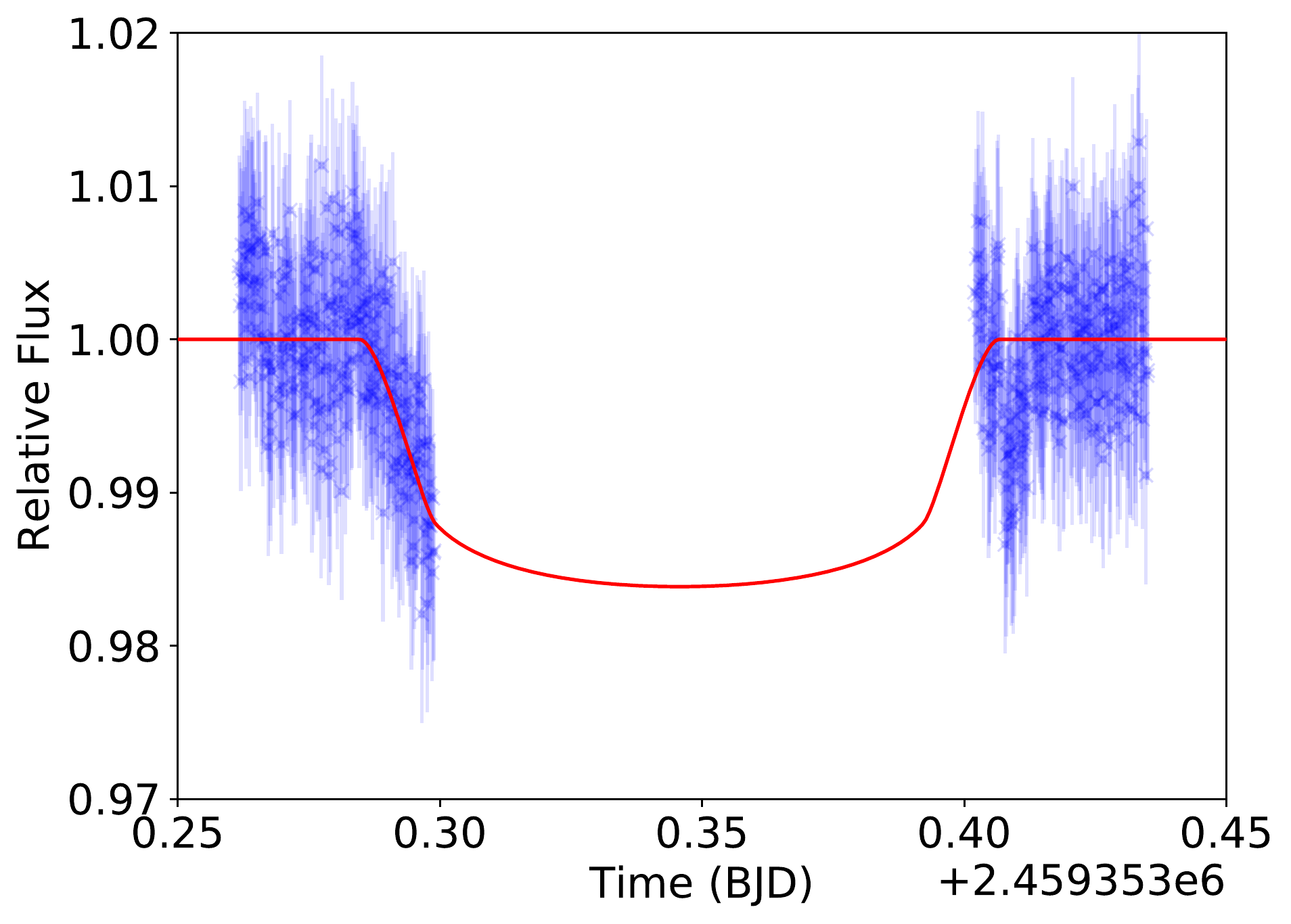} \\
 \includegraphics[scale=0.30]{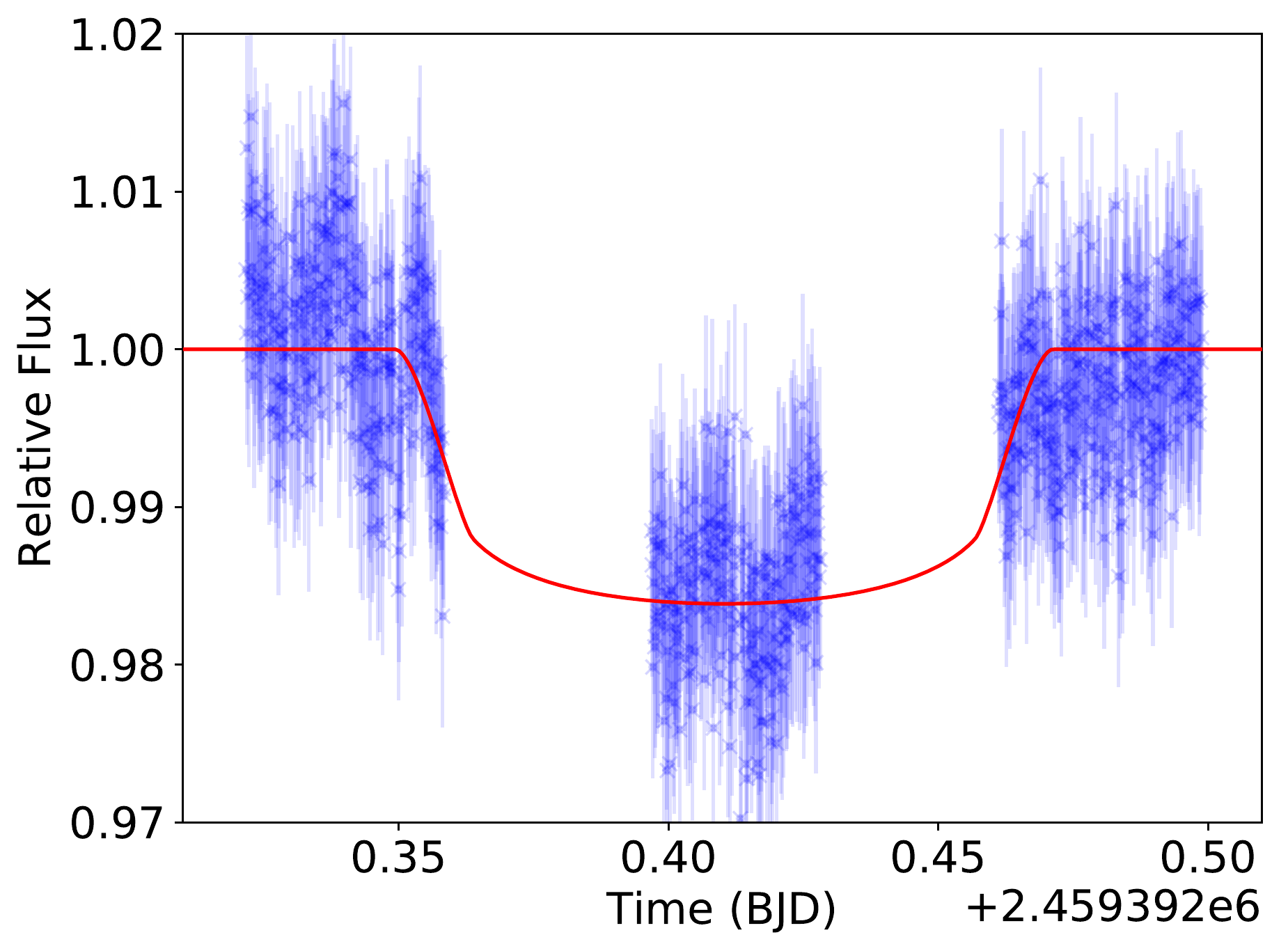}
 \includegraphics[scale=0.30]{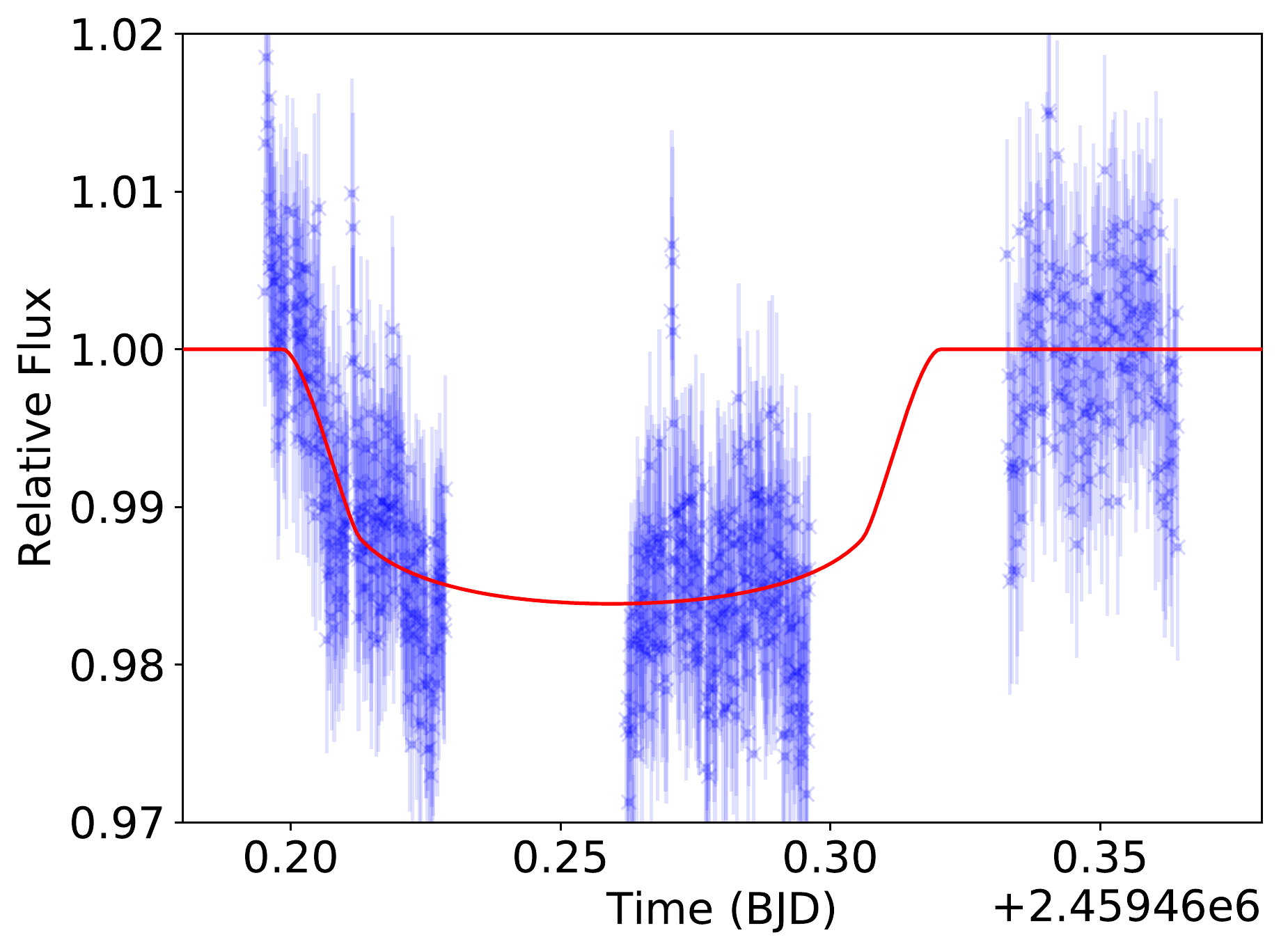}
 \includegraphics[scale=0.30]{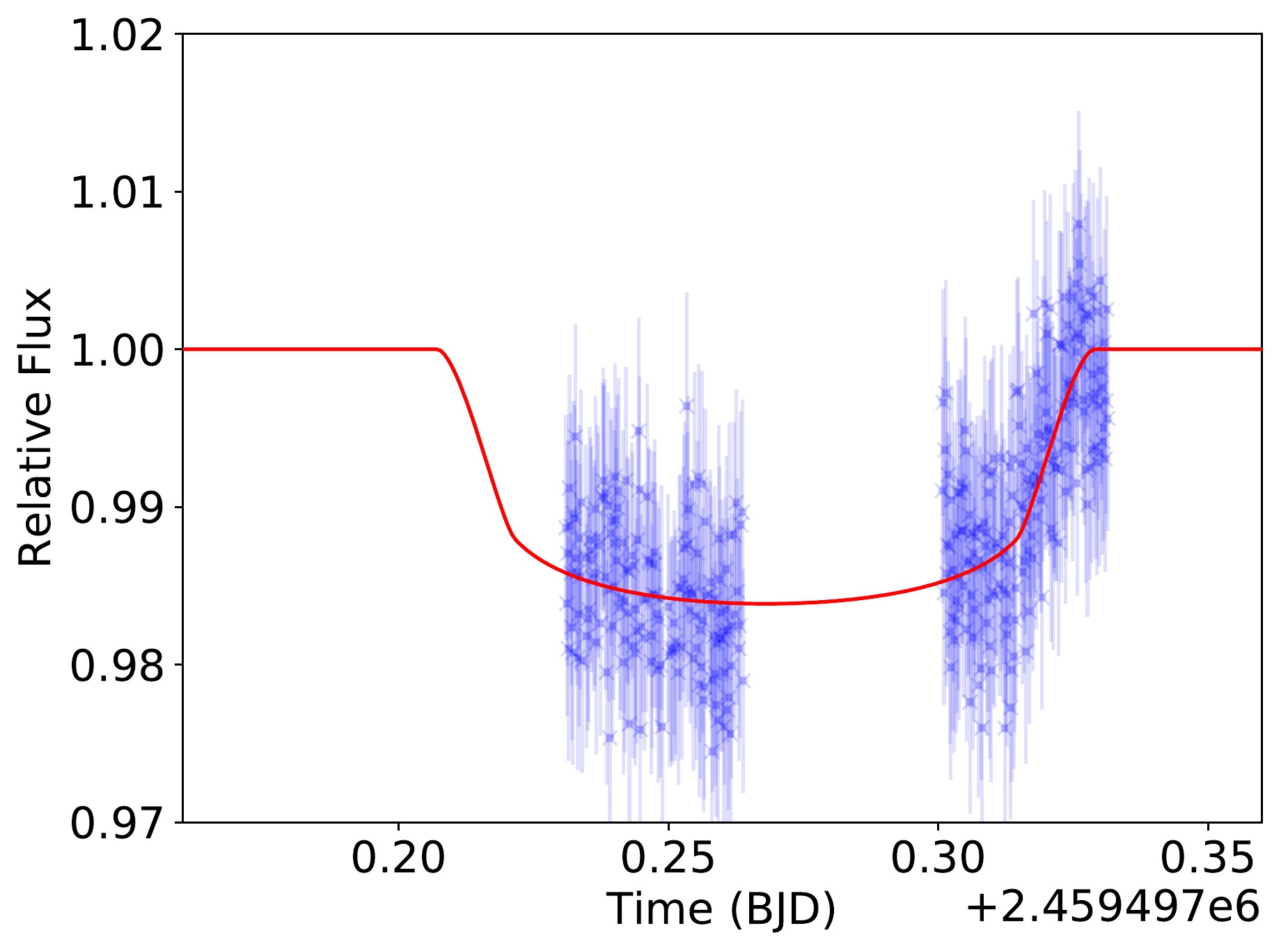}
 \caption{Normalized flux observations of of TOI-1516.01.  The red line is our fitted light curve. \label{fig:toi1516zoom} }
\end{figure*}

\begin{figure*}
 \includegraphics[scale=0.75]{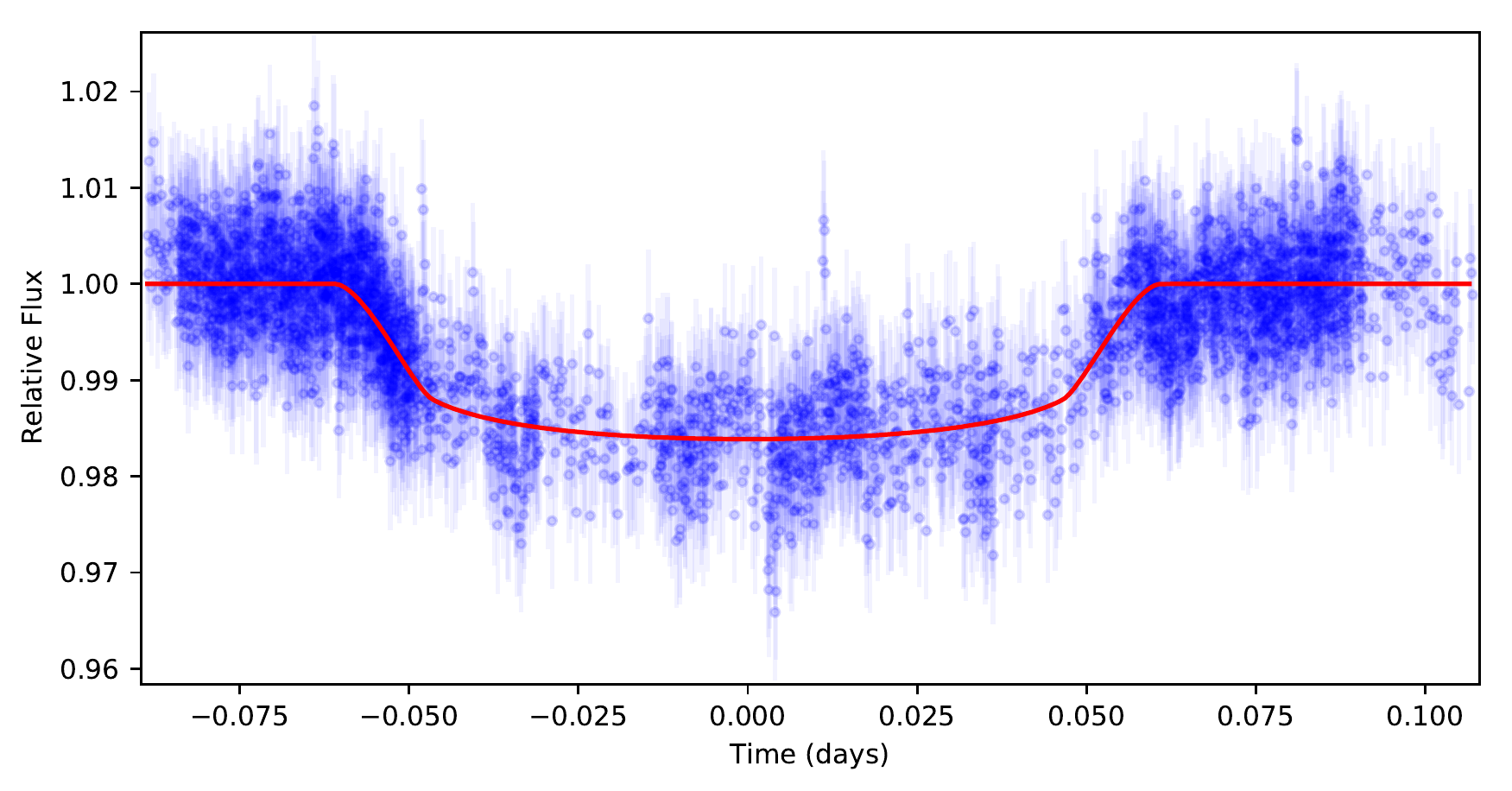}
 \hspace*{-0.17cm}\includegraphics[scale=0.75]{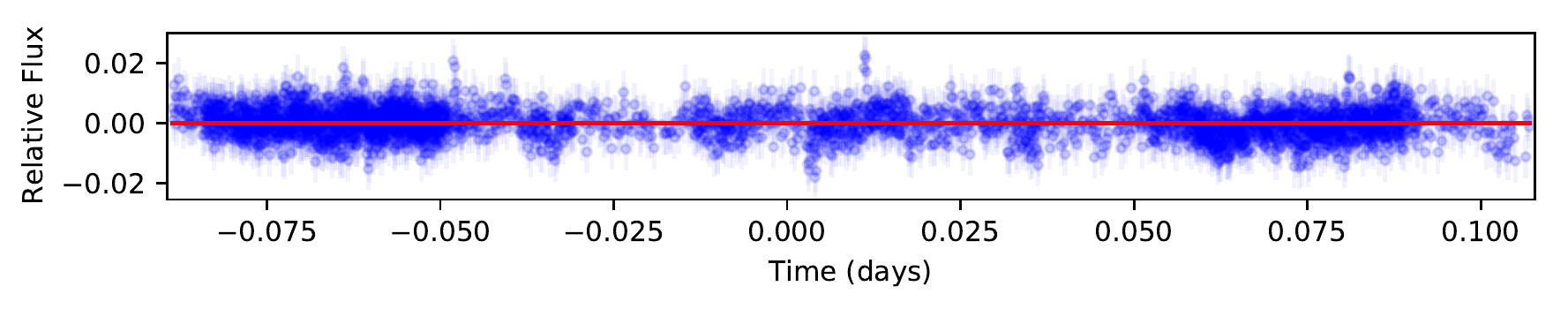}
 \caption{Phase-folded light curve and residuals of TOI-1516.01.  The red line is our fitted light curve.\label{fig:toi1516all} }
\end{figure*}

The individual transit events and folded curve are shown in figures \ref{fig:toi1516zoom} and \ref{fig:toi1516all} respectively.  Our results for TOI-1516.01, in table \ref{tab:allposts}, are consistent with what is already known of this hot Jupiter.  The planet has a radius 12\% of the host star's.  Four of the six transit events have good coverage of ingress, while a fifth has good egress coverage.  Based upon these transits, we find no evidence of TTVs of more than 1 minute.  The NEOSSat results for TOI-1516.01 are consistent with initial values reported by TESS (from the Exoplanet Transit Prediction Service \citep{nep2013}) as well as \citet{kabath2022}.

\subsection{TOI-2046}
TOI-2046.01 is a planetary candidate observed by TESS, observed in FFI (full-frame-image) in Sectors 18 and 19.  The host star is a $V=11.5$ F8 dwarf star. \citep{kabath2022}.  NEOSSat observations were taken on April 29, July 3, July 28, August 20, October 4, and November 22 of 2021, and one observation on June 9 of 2022.  The exposure time was 7 seconds with a 10 second cadence.  This planet was also validated by \citet{kabath2022} after the NEOSSat observations presented here were taken but before the writing of this paper.  The priors used for this target are shown in table~\ref{tab:allpriors}.
\begin{figure*}
 \includegraphics[scale=0.30]{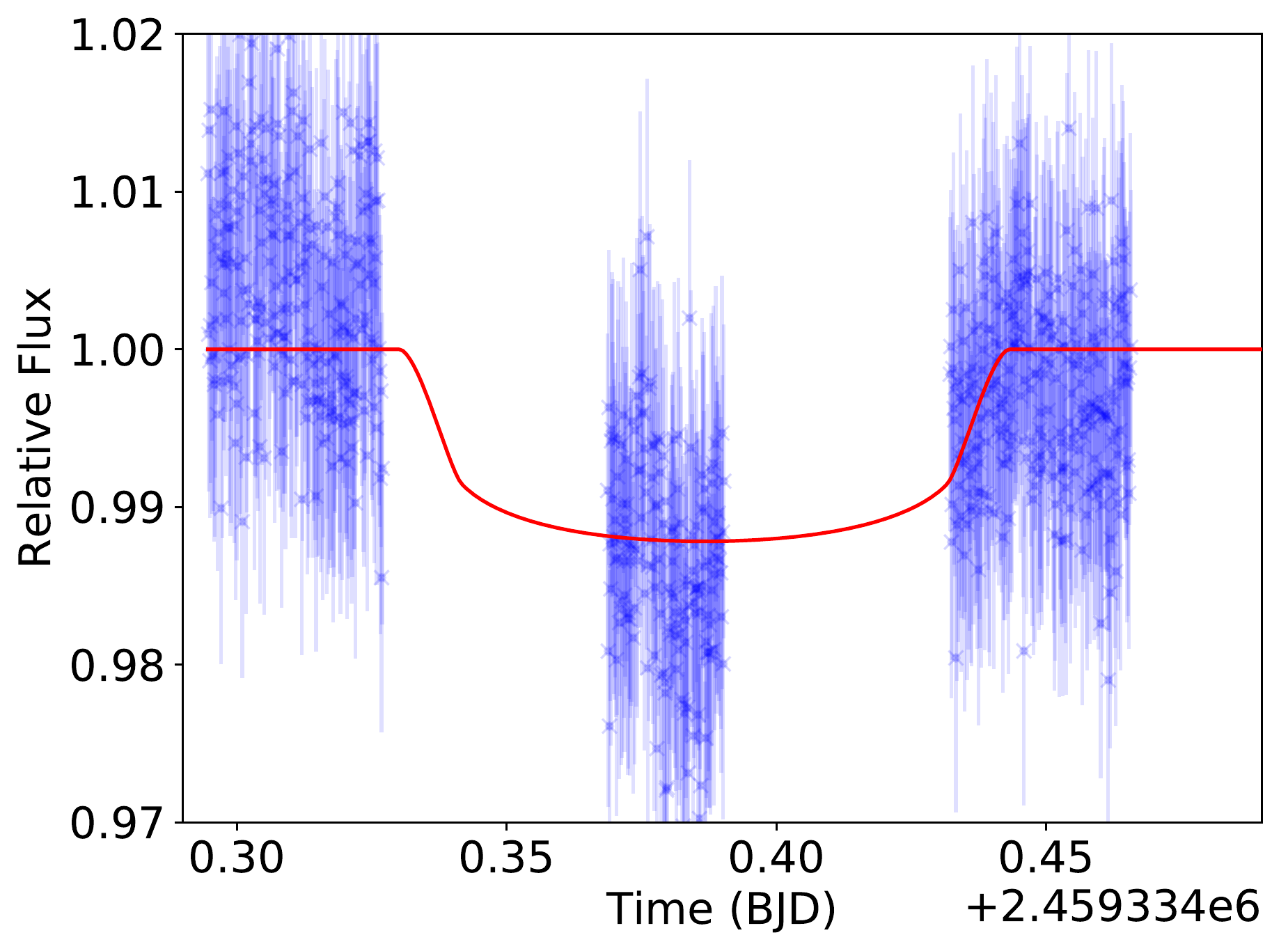}
 \includegraphics[scale=0.30]{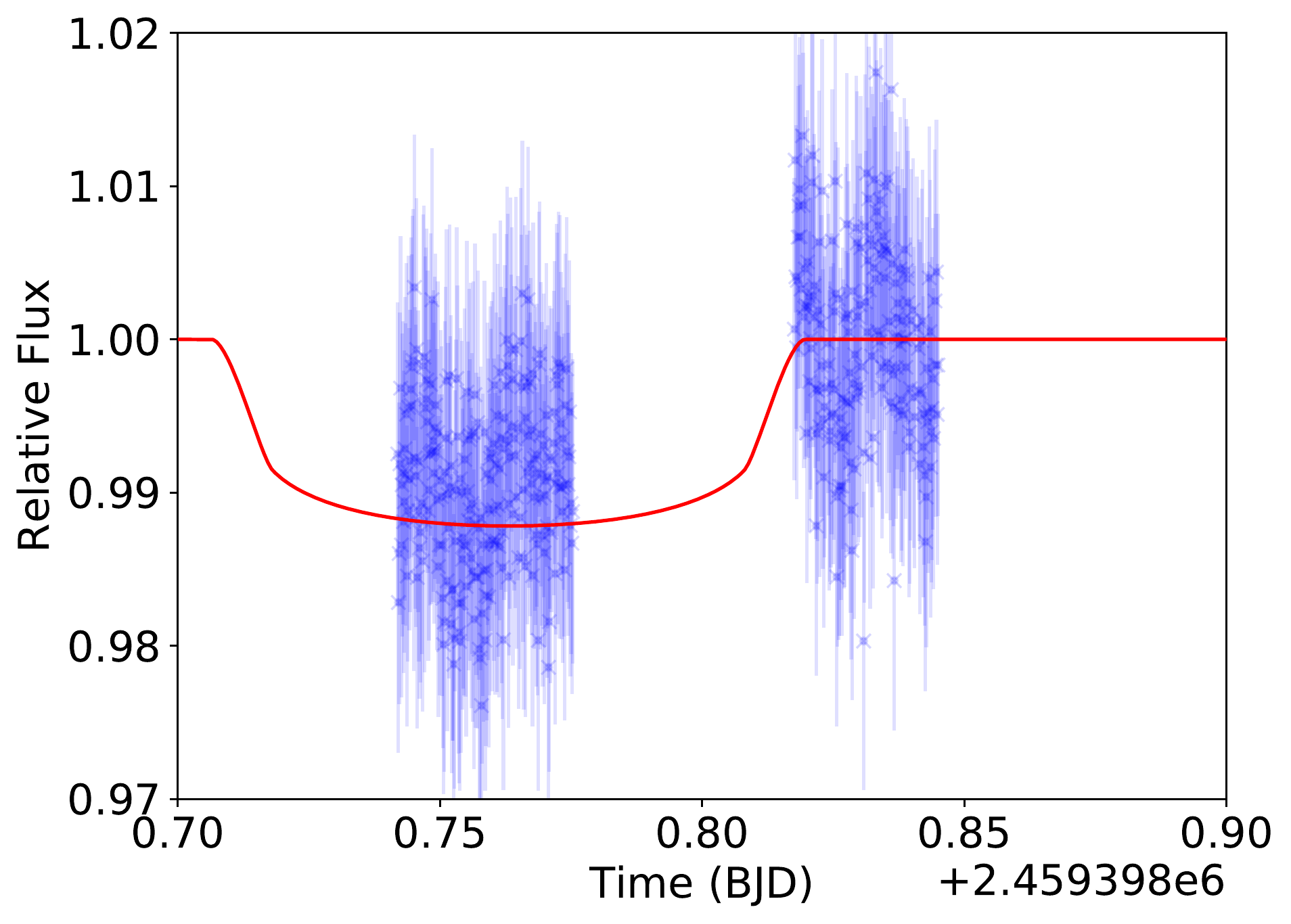}
 \includegraphics[scale=0.30]{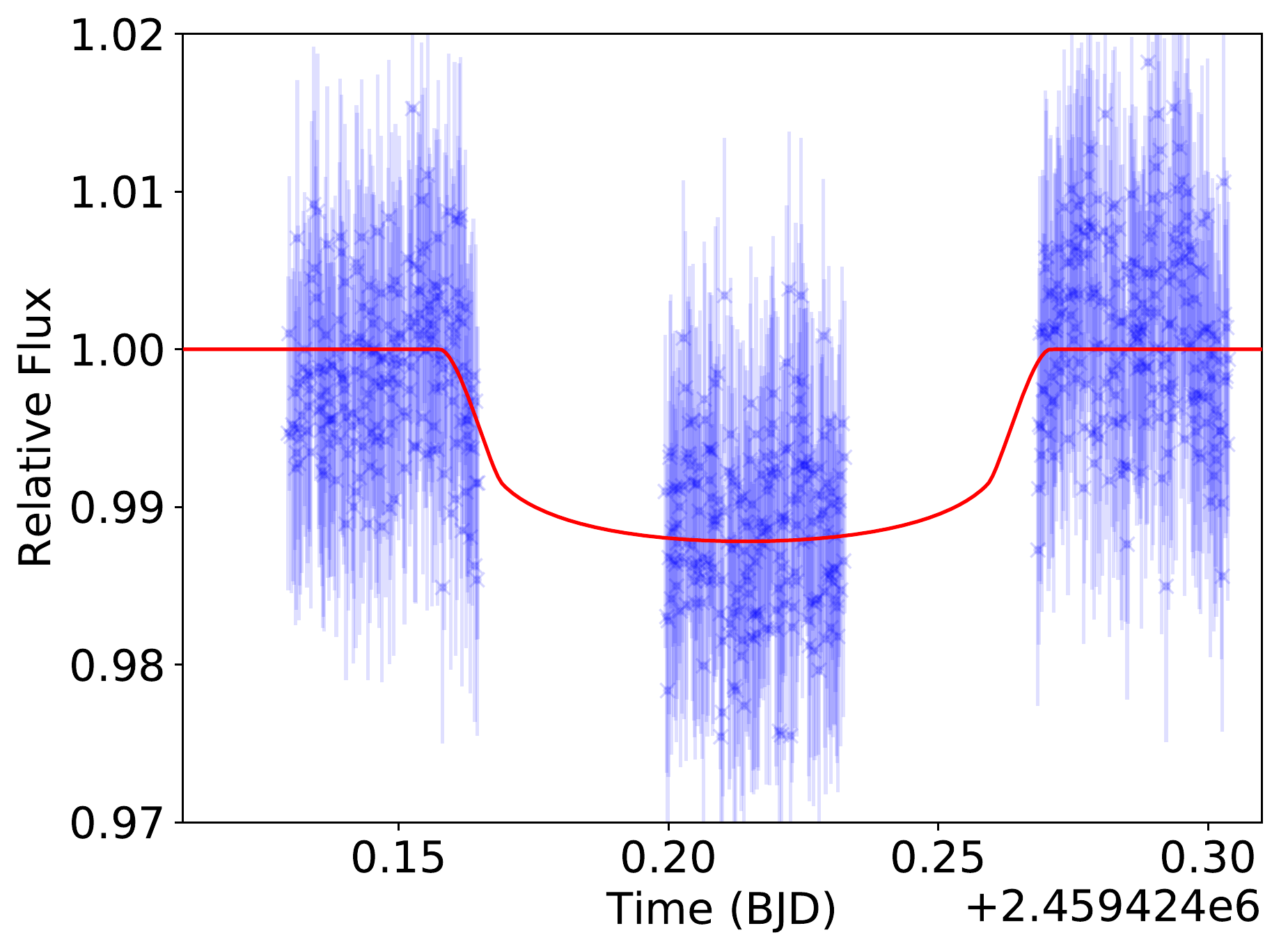} \\
 \includegraphics[scale=0.30]{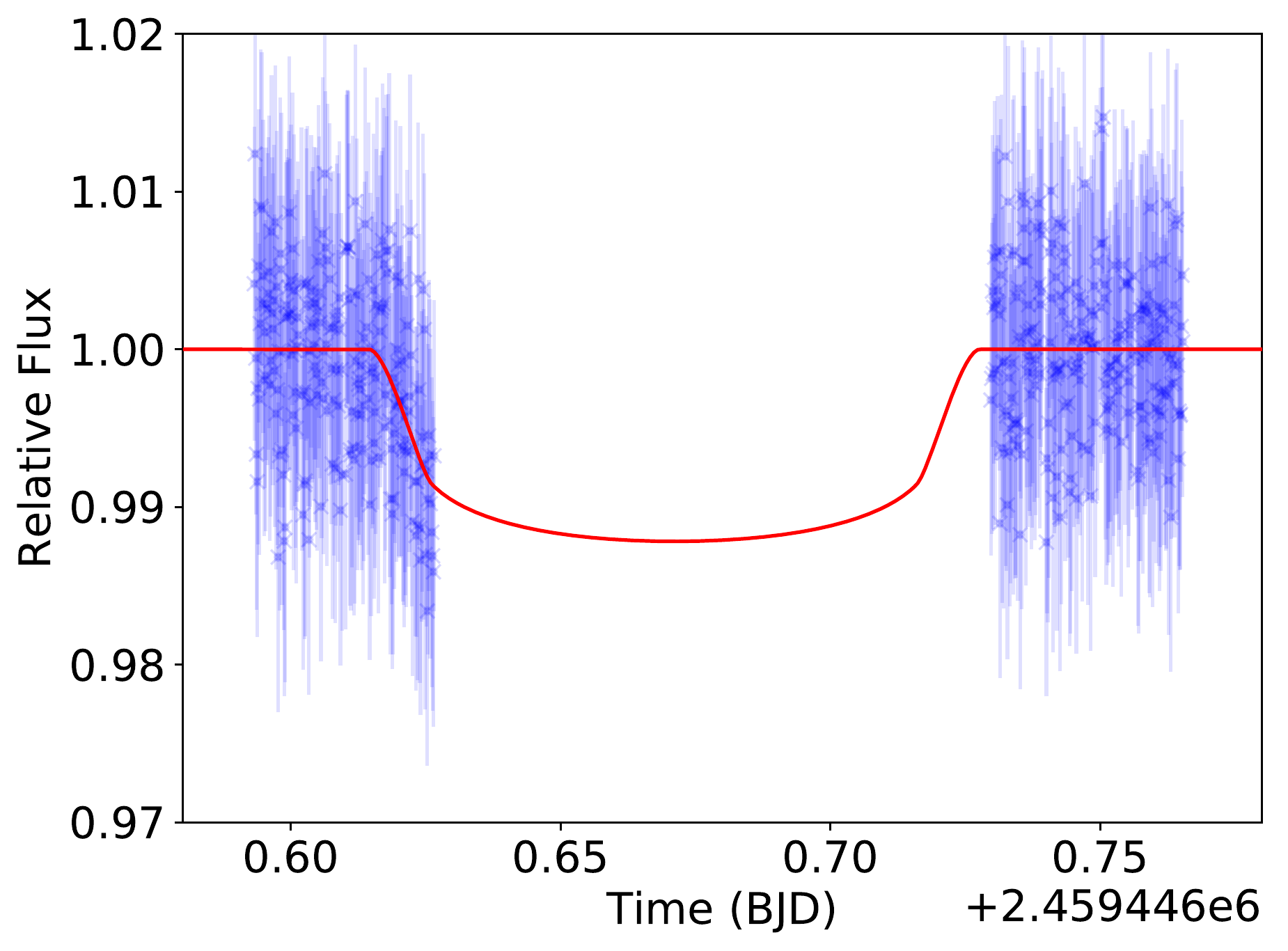}
 \includegraphics[scale=0.30]{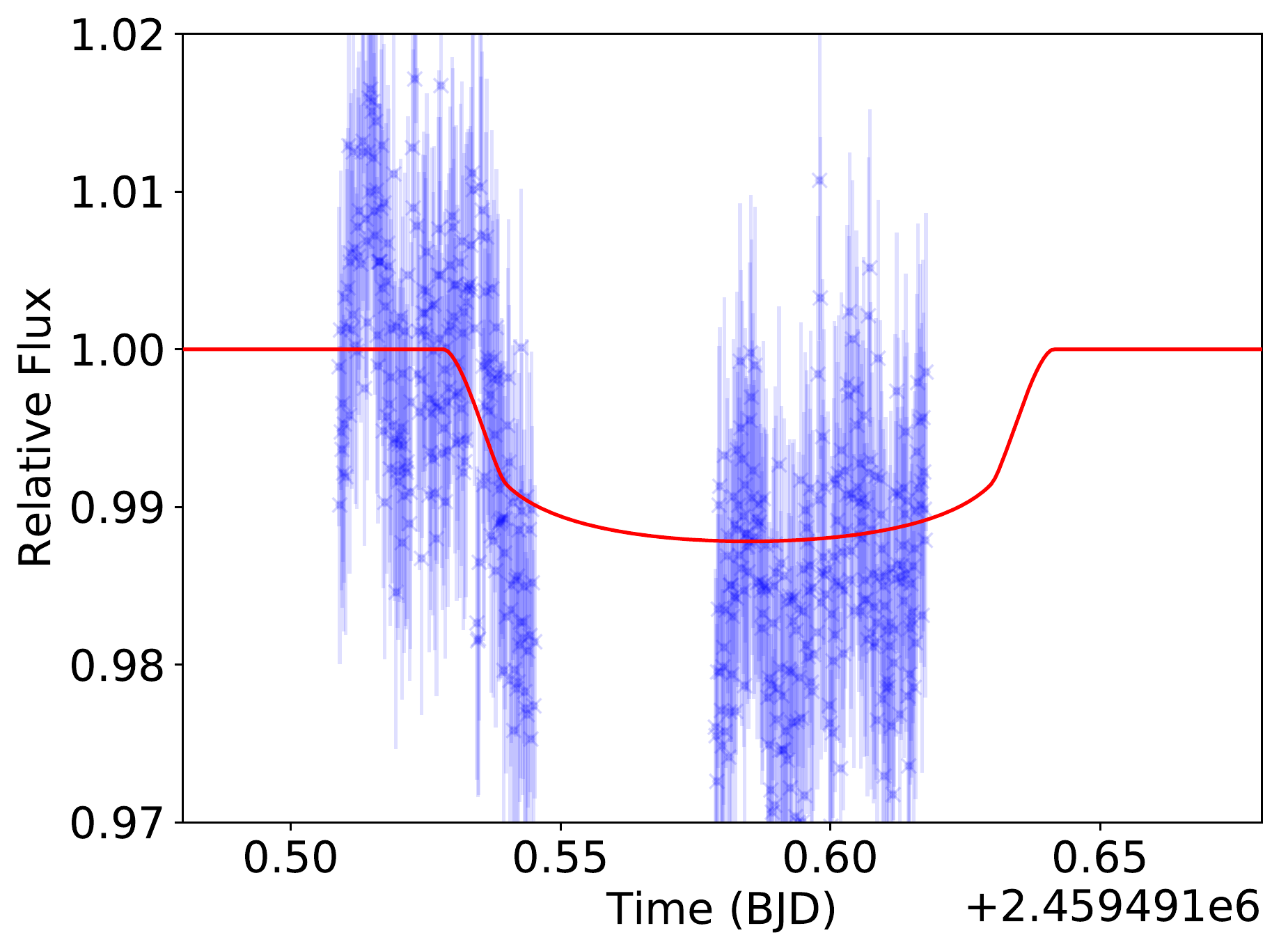} \\
 \includegraphics[scale=0.30]{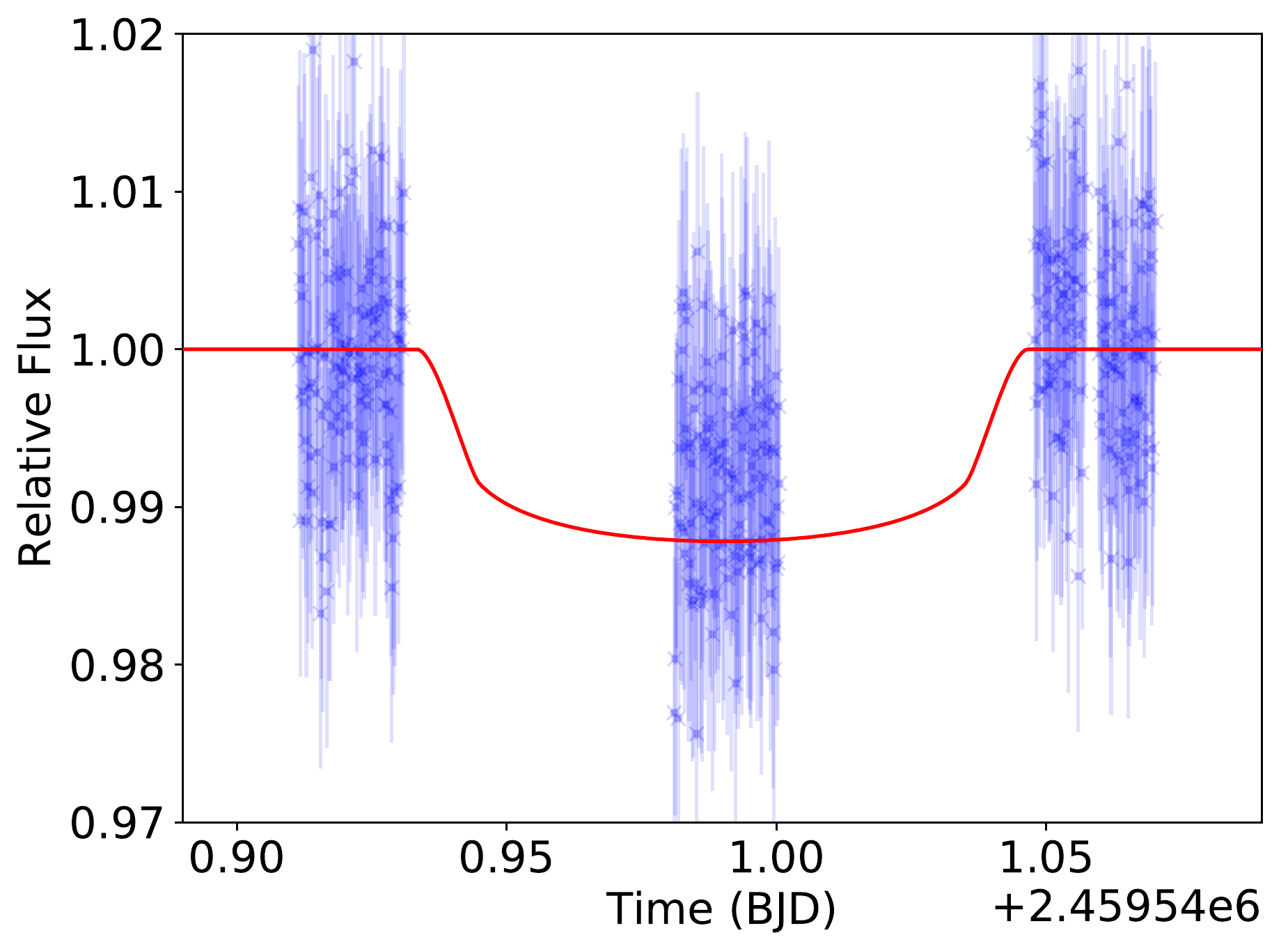}
 \includegraphics[scale=0.30]{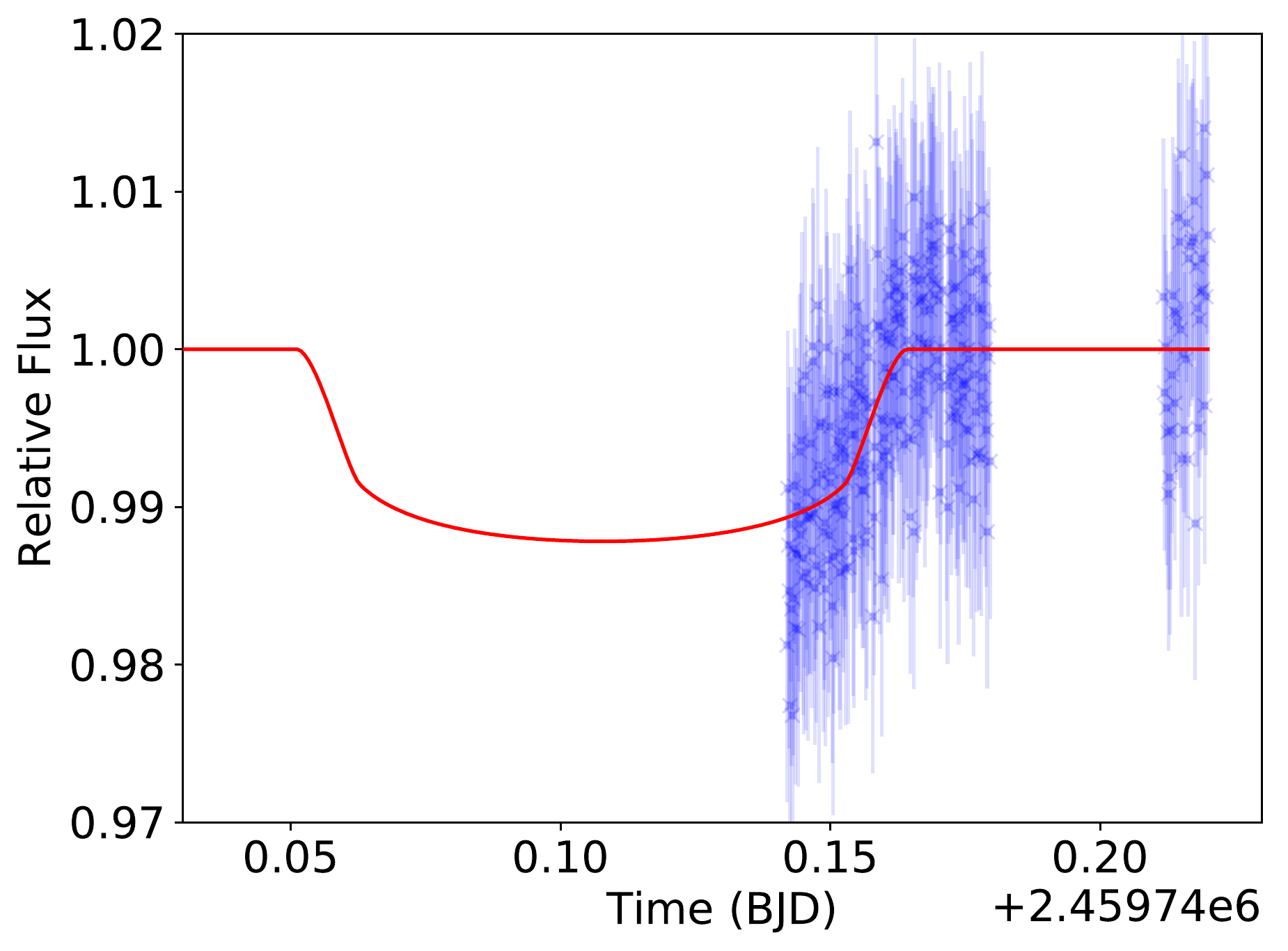}
 \caption{Normalized flux observations of of TOI-2046.01.  The red line is our fitted light curve.\label{fig:toi2046zoom} }
\end{figure*}

\begin{figure*}
 \includegraphics[scale=0.75]{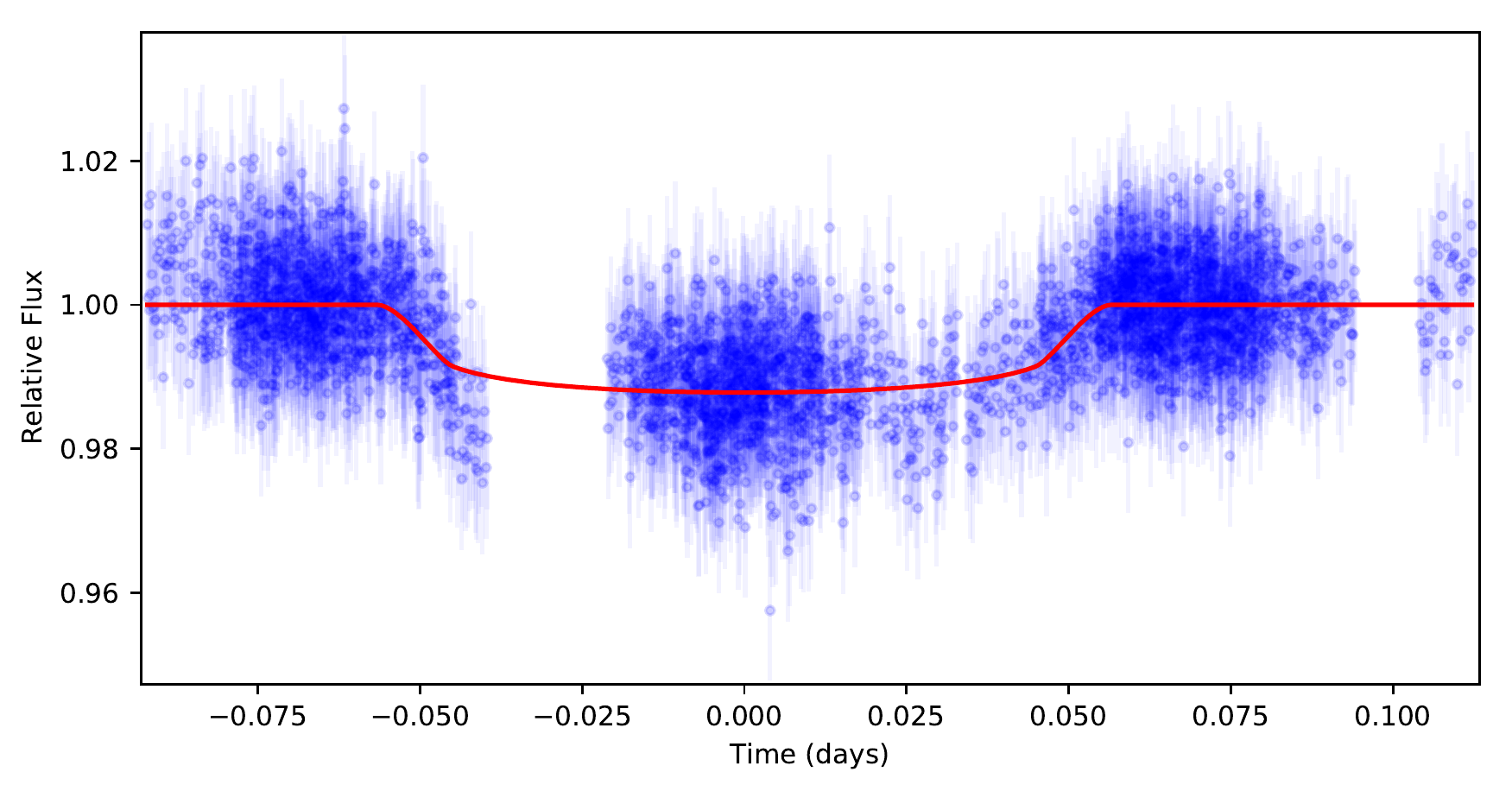}
 \hspace*{-0.31cm}\includegraphics[scale=0.75]{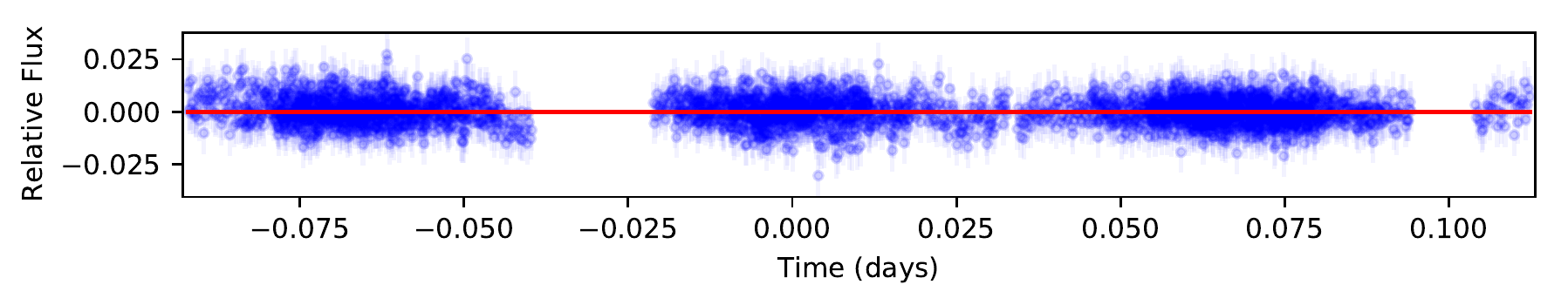}
 \caption{Phase-folded light curve and residuals of TOI-2046.01.  The red line is our fitted light curve.\label{fig:toi2046all} }
\end{figure*}

The individual transit events are shown in figure \ref{fig:toi2046zoom}, and the phase-folded light curve is shown in figure \ref{fig:toi2046all}.  Our results for TOI-2046.01, shown in table \ref{tab:allposts}, are consistent with the already-reported hot Jupiter nature of this system.  Only two of our seven observed events have coverage of ingress, and only one covers egress.  We see no indications of TTVs of larger than one minute.  The NEOSSat results for TOI-2046.01 are consistent with initial values reported by TESS (from the Exoplanet Transit Prediction Service \citep{nep2013}) as well as \citet{kabath2022}, with the exception of the planet:star radius.  While \citet{kabath2022} reports a value of 0.1213 $\pm$ 0.0017, we find a value of 0.101$\pm$0.002.  This difference is likely caused by some remaining systematic trends in the NEOSSat photometry.  

\begin{table*}
	\centering
	\caption{Posteriors and Computed Properties For Targets}
	\label{tab:allposts}
	\begin{tabular}{rcccl} 
	    Parameter & WASP-43 & TOI-1516 & TOI-2046 & Units\\
		\hline
		Planet Radius & 0.170 $\pm$ 0.018 & 0.1177 $\pm$ 0.0019 & 0.101 $\pm$ 0.002 & R$_{s}$ \\
		LDC 1 & 0.699 $\pm$ 0.029 & 0.348 $\pm$ 0.029 & 0.391 $\pm$ 0.028 &  \\
		LDC 2 & 0.060 $\pm$ 0.029 & 0.250 $\pm$ 0.029 & 0.245 $\pm$ 0.029 &  \\
		Central Time & 2459696.298550 $\pm$ 0.000757 & 2459392.41028 $\pm$ 0.00042 & 2459424.214180 $\pm$ 0.000854 & BJD \\
		Period &  0.81344697 $\pm$ 0.00002370 & 2.05603408 $\pm$ 0.00001256 & 1.49712564 $\pm$ 0.00000816 & days \\
		Semimajor Axis & 4.633 $\pm$ 0.251 & 5.834 $\pm$ 0.331 & 4.561 $\pm$ 0.252 & R$_{s}$ \\
		Inclination & 79.5 $\pm $1.1 & 87.02 $\pm $1.85 & 86.62 $\pm$ 2.35 & deg \\
		\hline
		Impact Parameter & 0.837 $\pm$ 0.098 & 0.302 $\pm$ 0.187 &  0.272 $\pm$ 0.187 &\\
		Total Duration ($t_{14}$) & 1.117 $\pm$ 0.187 & 2.8738 $\pm$ 0.2375 & 2.672 $\pm$ 0.221 & hours \\
		Stellar Density & 2.785 $\pm$ 0.465 & 0.901 $\pm$ 0.150 & 0.808 $\pm$ 0.136 & g/cm$^{3}$ \\
	\end{tabular}
\end{table*}

\section{Results and Analysis}
\subsection{Comparison to Other Results}
The purpose of this observing campaign was to demonstrate NEOSSat as a useful follow-up contributor to exoplanet science. However, for the particular systems targeted here, many of our results have been anticipated by the study of \cite{kabath2022} which examined two of our three targets, and which was published during the preparation of this manuscript.  As a result, though some of our results duplicate theirs, a comparison provides some insight into NEOSSat's performance vis-a-vis other techniques and systems currently in use.

\subsection{Basic system parameters}
In table \ref{tab:compare} we summarize the results of this paper together with established values from other sources.  Our results are largely consistent with other studies; our returned values are within the mutual error bars of the other sources. Thus NEOSSat is capable enough to provide basic exoplanetary system parameters with an accuracy comparable to that of other current observing programs.

\begin{table*}
	\centering
	\caption{Comparison of Basic System Parameters}
	\label{tab:compare}
	\begin{tabular}{rccccl} 
		\hline
		WASP-43 \\
	    Parameter & This work & \citet{hellier2011} & \citet{esposito2017} & \citet{patel2022} & Units\\
		\hline
		Plan. Rad. & 0.170$\pm$0.018 & - & 0.1588$\pm{0.0040}$ & 0.1615$\substack{+0.0017\\-0.0025}$ & R$_{s}$ \\ \\
		Period & 0.813447$\pm$0.000024 & 0.813475$\pm{0.000001}$ & 0.813473978$\pm{0.000000035}$ & 0.8134749$\substack{+0.0000009\\-0.0000010}$ & days \\ \\
		Semi. Axis & 4.633$\pm$0.251 & - & 4.97$\pm{0.14}$ & 4.72$\pm{0.05}$ & R$_{s}$ \\ \\
		Incl. & 79.5$\pm$1.1 & 82.6$\substack{+1.3\\-0.9}$ & 82.109$\pm{0.088}$& - & deg \\ \\
		Impact Par. & 0.837$\pm$0.098 & - & 0.689$\pm{0.013}$ & 0.698$\substack{+0.012\\-0.014}$ &  \\ \\
		Dur. ($t_{14}$) & 1.117$\pm$0.187 & 1.1592$\pm{0.0264}$ & 1.16$\pm{0.24}$ & - & hours \\ \\
		Stel. Dens. & 2.785$\pm$0.465 & 2.939$\substack{+0.971\\-0.545}$ & 2.43$\pm{0.14}$ & - & g/cm$^{3}$ \\
		\hline
		TOI-1516 \\
	    Parameter & This work & \citet{kabath2022} & \citet{TESSproject1516} & & Units\\
		\hline
		Plan. Rad. & 0.1177$\pm$0.0019 & 0.1224$\substack{+0.0005\\-0.0005}$ & 0.1212$\pm{0.0052}$& & R$_{s}$ \\ \\
		Period & 2.056034$\pm$0.000012 & 2.056014$\substack{+0.000002\\-0.000002}$ & 2.05603$\pm{0.00001}$ & & days \\ \\
		Semi. Axis & 5.834$\pm$0.331 & 6.22$\substack{+0.041\\-0.077}$ & - & & R$_{s}$ \\ \\
		Incl. & 87.02$\pm$1.85 & 90.0$\pm{0.4}$ & - & & deg \\ \\
		Impact Par. & 0.302$\pm$0.187 & 0.09$\substack{+0.10\\-0.07}$ & - & &\\ \\
		Dur. ($t_{14}$) & 2.8738$\pm$0.2375 & 2.826$\substack{+0.015\\-0.014}$ & 2.829$\pm{0.015}$& & hours \\ \\
		Stel. Dens. & 0.901$\pm$0.150 & 1.090$\pm{0.031}$ & 0.97417$\pm{0.219076}$ & & g/cm$^{3}$\\
		\hline
		TOI-2046 \\
	    Parameter & This work & \citet{kabath2022} & \citet{TESSproject2046}  & & Units\\
		\hline
		Plan. Rad. & 0.101$\pm$0.002 & 0.1213$\substack{+0.0017\\-0.0021}$ & 0.1354$\pm{0.0116}$ & & R$_{s}$ \\ \\
		Period & 1.49712564$\pm$0.0000082 & 1.4971842$\substack{+0.0000031\\-0.0000033}$ & 1.497 & & days \\ \\
		Semi. Axis & 4.561$\pm$0.252 & 4.75$\substack{+0.18\\-0.17}$ & - & & R$_{s}$ \\ \\
		Incl. & 86.62$\pm$2.35 & 83.6$\pm{0.9}$ & - & & deg \\ \\
		Impact Par. & 0.272$\pm$0.187 & 0.51$\substack{+0.06\\-0.07}$ & - & & \\ \\
		Dur. ($t_{14}$) & 2.672$\pm$0.221 & 2.410$\substack{+0.032\\-0.030}$ & 2.806$\pm{0.178}$ & & hours \\ \\
		Stel. Dens. & 0.808$\pm$0.136 & 0.890$\pm{0.098}$ & 0.835862$\pm{0.183146}$ & & g/cm$^{3}$ \\
		\hline
	\end{tabular}
\end{table*}

\subsection{Transit Ephemeris}
The ephemeris we find for these targets is consistent with those of TESS.  Using the central transit times from the TESS Project Candidate data \citep{TESSproject1516, TESSproject2046, TESSproject656} (depending on the target) as an initial start date, and the TESS computed period, we can estimate the difference between our computed central transit times and those expected from the TESS ephemeris.  For WASP-43b, TOI-1516.01, and TOI-2046.01 our central transit times are within 6.0, 3.7, and 0.9 minutes respectively (see table \ref{tab:refeph}).  Thus NEOSSat can produce central transit times consistent with other observatories.

Using the TESS results also gave us a longer baseline which enabled us to compute a new higher-precision ephemeris.  For TOI-1516.01, the resultant error on the period was reduced from $10^{-5}$ to $1.5\times10^{-6}$ days.  In the case of TOI-2046.01, the error was reduced from $10^{-4}$ (no error estimate is provided by the TESS Project, so we presumed the last digit) to $2.2\times10^{-6}$ days.  Our refined ephemeris for WASP-43b has a much higher error than the TESS project's, likely due to the very low error provided by the TESS Project.  This very low error is likely due to extending the baseline back to the initial WASP observations in 2010.  However, when combining our data with the TESS results, our refined ephemeris is within $4\times10^{-6}$ days (approximately 0.5 seconds) of other sources (\citet{esposito2017, patel2022}, among others). The updated ephemeris results are shown in table \ref{tab:refeph}.  Thus NEOSSat can be useful in refining orbital periods.

\begin{table*}
	\centering
	\caption{Refined Ephemeris For Targets}
	\label{tab:refeph}
	\begin{tabular}{rcccl} 
	     & WASP-43b & TOI-1516.01 & TOI-2046.01 & \\
		\hline
	    TESS T$_{c}$ & 2459279.797452 $\pm$ 0.000067 & 2458765.32531 $\pm$ 0.00019 & 2458792.39519 $\pm$ 0.00038  & BJD \\
        TESS Period & 0.813473629 $\pm$ 0.00000029 & 2.05603 $\pm$ 0.00001 & 1.4972 & days \\
	    TESS Expected T$_{c}$ & 2459696.295950 $\pm$ 0.000149 & 2459392.41446 $\pm$ 0.00305 & 2459424.21359 $\pm$ 0.00038 & BJD \\	
	    Our T$_{c}$ & 2459696.298550 $\pm$ 0.000757 & 2459392.410283 $\pm$ 0.000423 & 2459424.214180 $\pm$ 0.000854 & BJD \\
	    TESS Expected T$_{c}$ - Our T$_{c}$ & -0.002560 $\pm$ 0.000772 & 0.004177 $\pm$ 0.003079 & -0.000590 $\pm$ 0.000935 & days \\
		Baseline length & 416.5011 $\pm$ 0.000760 & 627.084973 $\pm$ 0.000464 & 631.81899 $\pm$ 0.000935 & days \\
		Number of Orbits & 512 & 305 & 422 & \\
		Resultant Period & 0.81347871 $\pm$ 0.00000148 & 2.05601631 $\pm$ 0.00000152 & 1.497201398 $\pm$ 0.00000222 & days \\
        \hline
	\end{tabular}
\end{table*}

\section{Summary}
In this study, we examined three hot Jupiters to demonstrate the capabilities of NEOSSat as a tool for exoplanet science.  We have demonstrated that NEOSSat can return exoplanetary parameters consistent with other dedicated exoplanet missions (such as TESS), even with incomplete transits events. Further, we used the results to improve the orbital ephemeris of these targets.  NEOSSat can be a useful tool for confirming and/or improving parameters.

However, there are some complications.  Because NEOSSat is generally limited to continuous viewing times of an hour, and transit durations are usually longer than this, finding brand new planets is a more difficult task.  A brand new planet could potentially have a period only known to an integer multiple due to the non-continuous nature of the coverage.  If certain stellar parameters can be known independently, then this period degeneracy could be reduced.  In such a case, multiple observations could be combined to produce a single curve with well-established parameters useful for followup science.

\newpage
\section{Acknowledgements}We thank the anonymous reviewer for their feedback that helped improve this work.  We also thank the NEOSSat Team at the Canadian Space Agency for their dedicated and enthusiastic help. This research used the facilities of the Canadian Astronomy Data Centre operated by the National Research Council of Canada with the support of the Canadian Space Agency. This work was funded in part by the Natural Sciences and Engineering Research Council of Canada Discovery Grants program (Grant no. RGPIN-2018-05659).

\section{Data Availability}
All data used in this paper comes from publicly available sources, including NASA's Exoplanet Archive at https://exoplanetarchive.ipac.caltech.edu).  Raw NEOSSat data is available online from the Canadian Astronomy Data Centre at https://www.cadc-ccda.hia-iha.nrc-cnrc.gc.ca/en/



\bibliographystyle{mnras}
\bibliography{neossat-3hj}





\bsp	
\label{lastpage}
\end{document}